\documentclass[12pt]{article}

\usepackage{amssymb,amsmath,amsfonts,eurosym,geometry,ulem,graphicx,caption,color,setspace,sectsty,comment,footmisc,caption,natbib,pdflscape,subfigure,array,hyperref}

\usepackage{algorithm}
\usepackage{algpseudocode}
\usepackage{booktabs}
\usepackage{amssymb}
\usepackage{amsmath}
\usepackage{bbm}
\usepackage{mathtools}

\DeclareMathOperator{\E}{\mathbb{E}}

\newcolumntype{L}[1]{>{\raggedright\let\newline\\arraybackslash\hspace{0pt}}m{#1}}
\newcolumntype{C}[1]{>{\centering\let\newline\\arraybackslash\hspace{0pt}}m{#1}}
\newcolumntype{R}[1]{>{\raggedleft\let\newline\\arraybackslash\hspace{0pt}}m{#1}}

\begin{document}

\begin{titlepage}
\title{Simple models predict behavior at least as well as behavioral scientists}
\author{Dillon Bowen\thanks{Wharton School of Business, dsbowen@wharton.upenn.edu}}
\date{\today}
\maketitle
\begin{abstract}
\noindent How accurately can behavioral scientists predict behavior? To answer this question, we analyzed data from five studies in which 640 professional behavioral scientists predicted the results of one or more behavioral science experiments. We compared the behavioral scientists' predictions to random chance, linear models, and simple heuristics like ``behavioral interventions have no effect" and ``all published psychology research is false." We find that behavioral scientists are consistently no better than - and often worse than - these simple heuristics and models. Behavioral scientists' predictions are not only noisy but also biased. They systematically overestimate how well behavioral science ``works": overestimating the effectiveness of behavioral interventions, the impact of psychological phenomena like time discounting, and the replicability of published psychology research \footnote{Code available here: \url{https://gitlab.com/dsbowen/publication-bias}}\\
\vspace{0in}

\bigskip
\end{abstract}
\setcounter{page}{0}
\thispagestyle{empty}
\end{titlepage}
\pagebreak \newpage

\doublespacing

\section{Introduction}

Behavioral scientists' predictions regularly inform academia, public policy, and business decisions. Academic researchers decide which projects to pursue based on which hypotheses seem most plausible. Nudge units test and implement interventions they expect will be most effective. Businesses regularly consult marketing and management experts for advice.

The underlying assumption in these instances is that behavioral scientists' predictions are accurate. In general, we expect experts in any domain to be able to make accurate predictions about their domain of expertise \citep{tetlock2009expert}. We should have especially high expectations of scientific experts, given that prediction is a fundamental function of science \citep{kuhn20119}. We might even argue that accurate predictions are necessary for the credibility of a scientific discipline \citep{tetlock2009expert, tetlock2016superforecasting}.

These considerations raise a fundamental question: \textit{How accurately can behavioral scientists predict behavior?}

To answer this question, we examined five studies in which behavioral scientists predicted the results of one or more behavioral science experiments \citep{milkman2021megastudies, milkman2022680, dellavigna2018predicting, dellavigna2022rcts, dreber2015using}. The goals of these studies included testing methods for aggregating predictions \citep{dreber2015using}, comparing subgroups of behavioral scientists \citep{dellavigna2018predicting, dellavigna2022rcts}, and comparing behavioral scientists to non-scientists and scientists in other domains \citep{milkman2021megastudies}. However, they did not compare the behavioral scientists' predictions to simple benchmarks (e.g., random chance) or heuristics (e.g., behavioral interventions have no effect). Such comparisons are critical considering research demonstrating that simple models often perform at least as well as experts in political science \citep{mellers2014psychological, tetlock2009expert, tetlock2016superforecasting}, clinical diagnosis \citep{dawes1974linear, dawes1989clinical, garb1989clinical}, and other domains \citep{dawes1979robust, dawes1974linear}.

\textbf{Exercise study.} The first experiment used 53 behavioral nudges to encourage 24-Hour Fitness customers to exercise more \citep{milkman2021megastudies}. The researchers measured the nudges' effectiveness as the increase in average weekly gym visits during a four-week intervention compared to a control condition. They then asked 90 practitioners from behavioral science companies to predict how effective each nudge would be.

We compared the behavioral scientists' predictions to a null model that predicted that none of the behavioral nudges would increase exercise. Specifically, the null model predicted that all nudges would increase exercise by zero gym visits per week compared to the control condition.

\textbf{Flu study.} The second experiment used 22 text-message treatments to encourage Walmart customers to get a flu vaccine \citep{milkman2022680}. The treatments varied the text messages' phrasing and the number of messages sent. The researchers measured a treatment's effectiveness as the increase in vaccination rates (number of people per hundred) compared to a control condition in which they did not text customers. They then asked 24 professors and graduate students, most affiliated with top-10 business schools, to predict how effective each treatment would be.

As in the exercise study, we compared the behavioral scientists to a null model that predicted that none of the text-message treatments would increase vaccination rates. Specifically, the null model predicted that all text-message treatments would increase vaccination rates by zero people per hundred compared to the control condition.

\textbf{RCT study.} The third study assembled data from 126 randomized control trials (RCTs) from two of the largest Nudge Units in the United States \citep{dellavigna2022rcts}. The researchers measured the RCTs' effectiveness as the percentage point increase in adopting a target behavior compared to a control condition. They then asked 237 behavioral scientists from academia, non-profits, government agencies, and nudge units to estimate the effectiveness of 14 randomly selected RCTs.

We compared the behavioral scientists' predictions to a null model that predicted that none of the nudges would increase adoption. Specifically, the null model predicted that all nudges would increase the adoption of the target behavior by zero percentage points compared to the control condition.

\textbf{Effort study.} The fourth experiment measured how much effort participants exerted in a key-pressing task in 18 experimental conditions \citep{dellavigna2018motivates}. Participants scored points for alternating between pressing ``a" and ``b" for 10 minutes (participants earned one point each time they pressed ``a" then ``b"). The experimental conditions were monetary and non-monetary incentives to score points, such as piece-rate payments, time-delayed payments, and peer comparisons. The researchers measured the effort participants exerted in each condition as the number of points they scored. They then recruited 213 academic economists to predict their results \citep{dellavigna2018predicting}. The researchers showed the economists the results of three experimental conditions: how much effort participants exerted when the researchers did not pay them, when they paid them 1 cent per 100 points, and when they paid them 10 cents per 100 points. The economists then predicted the results of the other 15 conditions.

\begin{figure}
    \centering
    \includegraphics[scale=.7]{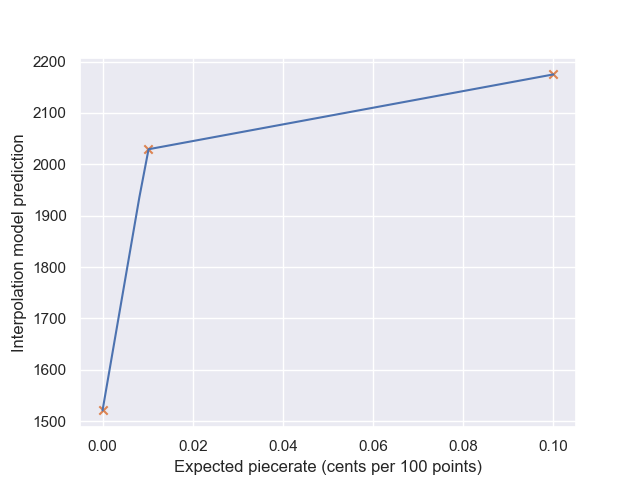}
    \caption{Linear interpolation model for the effort study. The three orange x's are the three conditions for which the economists saw the results (no payment, 1-cent piece-rate, and 10-cent piece-rate).}
    \label{fig:effort_model}
\end{figure}

We compared the economists' predictions to a simple linear interpolation between the three conditions the economists saw (see Figure \ref{fig:effort_model}). This model makes the simplifying assumption that participants only care about their expected piece-rate payment. Importantly, it assumes that participants do not exhibit motivational crowding out, time discounting, risk aversion, framing effects, or other psychological tendencies commonly studied by behavioral economists.

This linear interpolation model can be ``selfish" or ``altruistic." In two experimental conditions, participants earned a piece-rate payment on behalf of the Red Cross (e.g., the researchers donated 1 cent to the Red Cross for every 100 points the participant scored). The selfish version of the linear interpolation model assumes that participants do not care about the Red Cross. The altruistic version assumes that participants are just as motivated to earn money for the Red Cross as they are to earn money for themselves.

\textbf{Reproducibility study.} The fifth study attempted to reproduce 100 psychology studies published in top academic journals \citep{open2015estimating}. The researchers defined a replication as successful if it obtained a p-value of less than .05 and the estimated effect's direction matched the original experiment. They then asked 76 psychology professors and graduate students to predict how likely 44 of the studies were to replicate successfully \citep{dreber2015using}. We compared the psychologists' predictions to three benchmarks:

\begin{enumerate}
    \item \textit{Null model.} All published psychology research is false (i.e., the null hypothesis is always true). Therefore, the probability of a successful replication is 2.5\% (5\% chance of $p<.05$, half the time the direction of the estimated effects in the original and replication experiments will match).
    \item \textit{Random chance.} There is a 50\% chance each study will replicate.
    \item \textit{Linear regression model.} This model predicts the estimated effect in the replication study using the estimated effect in the original study. It then uses this prediction to estimate a study's probability of replicating.
\end{enumerate}

\subsection{Dependent measures}

We scored predictions using squared error for the exercise, flu, RCT, and effort studies. Because the reproducibility study used a binary outcome variable (an indicator that the study successfully replicated), we scored predictions using Brier scores. We define \textit{risk} as the expected score. Our primary dependent measure was \textit{comparative risk}, the expected difference between the behavioral scientists' and models' scores. We also analyze \textit{bias}, the expected difference between the behavioral scientists' predictions and the treatment effects (or, for the reproducibility study, the probability of replicating).

To measure bias in the effort study, we first grouped the experimental conditions into eight categories described in the effort study paper. The categories group conditions that measure the same psychological phenomenon (e.g., social preferences). Then, we reverse coded some of the categories. For example, economic theory suggests that participants should exert less effort when payments are time-delayed. Therefore, a behavioral scientist overestimates the effect of time discounting when she underestimates the number of points scored in the time-delayed category. Finally, we measured bias as the expected difference between the behavioral scientists' predictions and the number of points scored in each category\footnote{The exception to this is the framing effects category. We measure framing effects as the difference between the loss-framed and gain-framed conditions. Behavioral scientists are biased if they overestimate the difference between these conditions. See Section \ref{section:methods} for a formal description of how we computed bias for the framing effects category.}.

Throughout our analysis, we compare the behavioral scientists and models to an ``oracle" to estimate the irreducible uncertainty. The oracle makes the best possible prediction given the experiment results. The oracle's risk is positive because the oracle only knows the experimental results, not the true effects. For example, because the oracle does not know the true effects of the exercise study, its expected squared error will be positive.

\section{Results}

\begin{table}
\centering
\caption{Risk for behavioral scientists and models.}
\begin{tabular}{llrr}
Study & Predictor & Mean & 95\% CI \\
\midrule
Exercise & Scientist & 6.02 & (5.19, 6.93) \\
 & Null & 0.03 & (0.01, 0.05) \\
 & Oracle & 0.00 & (0.00, 0.01) \\
Flu & Scientist & 64.24 & (14.80, 184.52) \\
 & Null & 4.80 & (2.76, 7.28) \\
 & Oracle & 0.11 & (0.03, 0.38) \\
RCT & Scientist & 54.59 & (27.97, 101.94) \\
 & Null & 7.48 & (2.24, 17.04) \\
 & Oracle & 1.13 & (0.18, 4.70) \\
Effort & Scientist & 49884 & (38858, 64352) \\
 & Selfish & 49865 & (26356, 80066) \\
 & Altruistic & 34639 & (18785, 55836) \\
 & Oracle & 727 & (232, 1636) \\
Reproducibility & Scientist & 0.31 & (0.28, 0.35) \\
 & Null & 0.31 & (0.20, 0.43) \\
 & Random & 0.25 & (0.25, 0.25) \\
 & Linear regression & 0.23 & (0.18, 0.28) \\
 & Oracle & 0.16 & (0.12, 0.21) \\
\bottomrule
\end{tabular}
\label{table:risk}
\end{table}

\begin{table}
\centering
\caption{Bias.}
\begin{tabular}{llrr}
Study & Units & Mean & 95\% CI \\
\midrule
Exercise & Weekly gym visits & 2.32 & (2.15, 2.50) \\
Flu & People per 100 & 2.41 & (0.33, 5.90) \\
RCT & Percentage points & 3.37 & (1.75, 5.31) \\
Effort & Points & 66.90 & (23.35, 126.63) \\
Reproducibility & Pr. replication & 0.22 & (0.08, 0.33) \\
\bottomrule
\end{tabular}
\label{table:bias}
\end{table}

\textbf{Exercise study.} The behavioral scientists' exercise study predictions were significantly worse than the null model (see Table \ref{table:risk}). That is, the comparative risk was significantly positive ($M=6.00, P < .001$). Behavioral scientists estimated that the average behavioral nudge would increase exercise by 2.5 gym visits every week. The study results suggest that the average nudge increased exercise by only one gym visit every six weeks. This bias was statistically significant ($P < .001$, see Table \ref{table:bias}). Even after applying a correction for multiple hypothesis testing \citep{romano2005stepwise, kitagawa2018should, mogstad2020inference}, behavioral scientists significantly overestimated the effectiveness of every treatment in the study.

\textbf{Flu study.} The behavioral scientists' flu study predictions were also significantly worse than the null model ($M=59.44, P < .001$). Behavioral scientists estimated that the average text-message treatment would increase vaccination rates by 4.5 people per hundred. The study results suggest that the average treatment increased vaccination rates by only 2.1 people per hundred.  This bias was statistically significant ($P=.015$).

\textbf{RCT study.} The null model also outperformed behavioral scientists in the RCT study ($M=47.11, P < .001$). Behavioral scientists estimated that the average nudge would increase adoption of the target behavior by 4.98 percentage points. The study results indicate that the average nudge increased adoption by only 1.61 percentage points. This bias was statistically significant ($P=.001$).

The RCT prediction study dataset also included information about the behavioral scientists' experience with RCTs (i.e., the number of nudge RCTs they had run). The authors grouped the behavioral scientists into three categories: those with no experience running RCTs (novices), those who had run 1-5 RCTs (moderately experienced), and those who had run more than five RCTs (most experienced). They observed that more experienced behavioral scientists were less biased based on summary statistics.

\begin{figure}
    \centering
    \includegraphics[scale=.7]{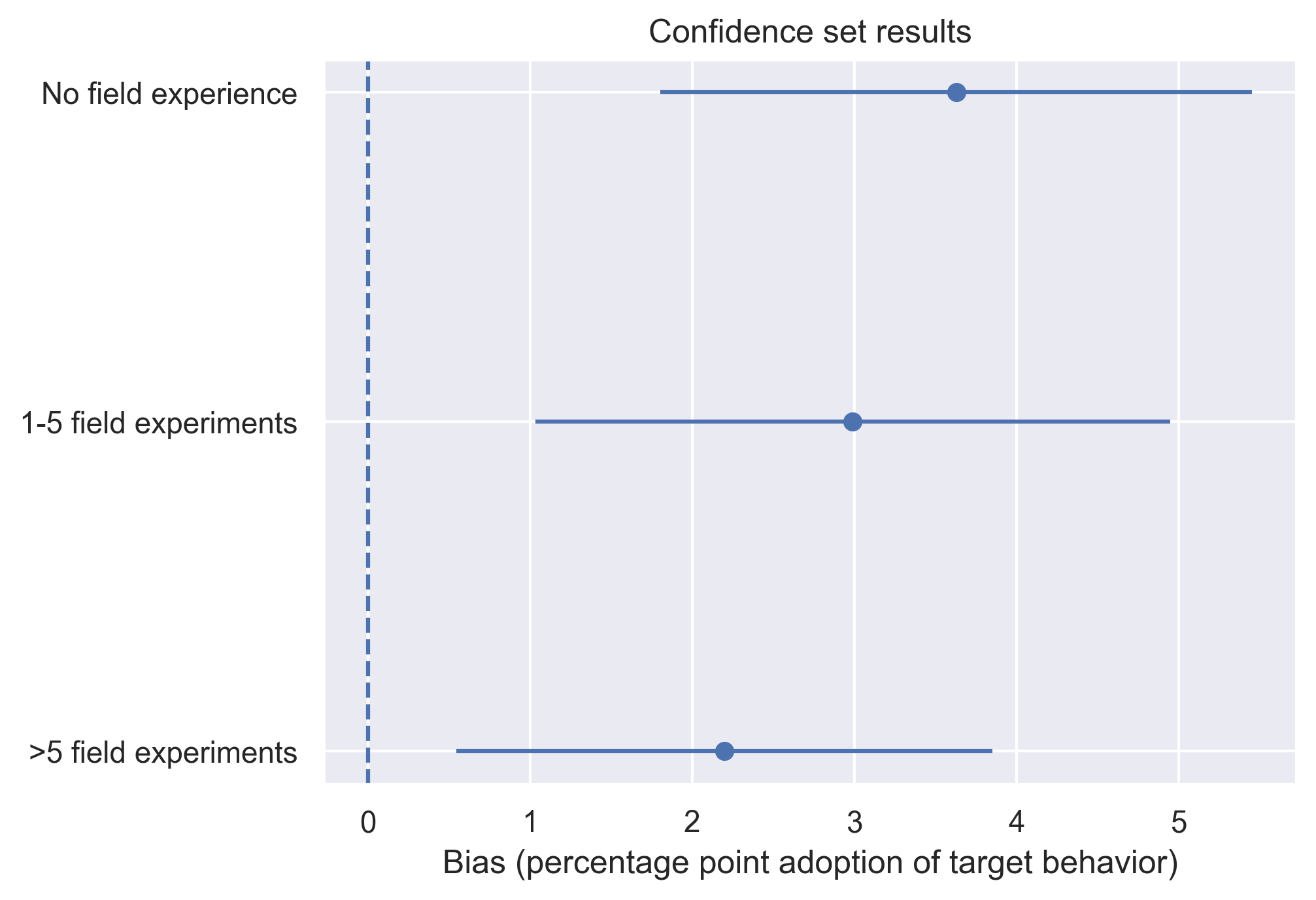}
    \caption{Bias for the novice, moderaly experienced, and most experienced behavioral scienitists. Error bars show simultaneous 95\% confidence intervals.}
    \label{fig:bias_by_experience}
\end{figure}

We formally analyzed the relationship between experience and bias in the RCT study (see Figure \ref{fig:bias_by_experience}). Consistent with the RCT study's conclusion, we estimate that the most experienced behavioral scientists are 0.65 percentage points less biased than the moderately experienced behavioral scientists and 1.44 percentage points less biased than the novices. However, pairwise hypothesis tests suggest that none of the experience groups significantly differ from any other. Additionally, even the most experienced behavioral scientists overestimate nudges' effectiveness by 2.20 percentage points ($P=.005$ after correcting for multiple hypothesis testing).

\textbf{Effort study.} The economists' effort study predictions were worse than the selfish and altruistic linear interpolation models. However, the differences are not statistically significant ($M=19.38, P=1.000$ for the selfish model, $M=15245.30, P=.108$ for the altruistic model, p-values adjusted for multiple hypothesis testing).

\begin{figure}
    \centering
    \includegraphics[scale=.7]{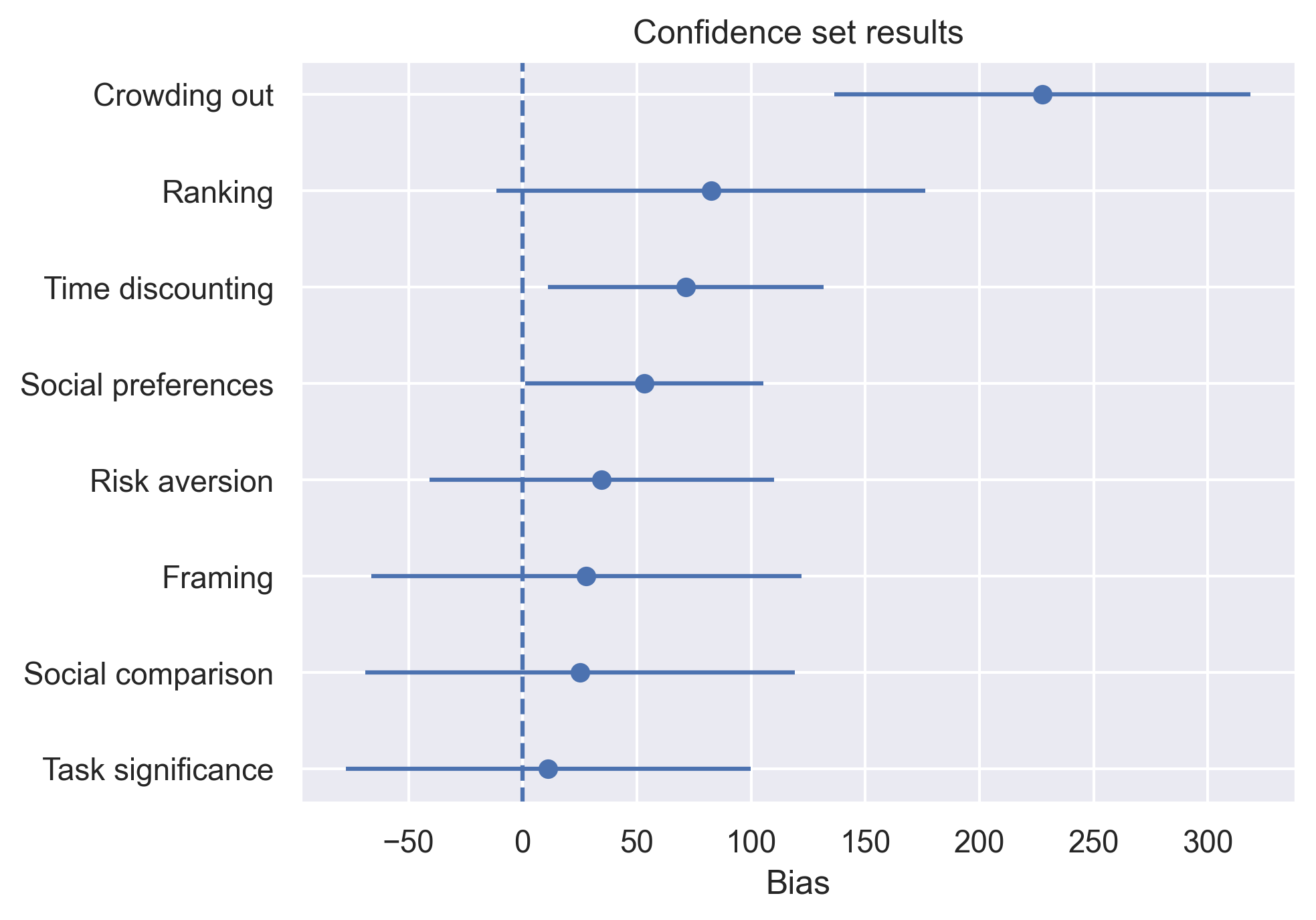}
    \caption{Bias for the effort study. Error bars show simultaneous 95\% confidence intervals.}
    \label{fig:bias}
\end{figure}

Economists overestimated the impact of psychological phenomena by 67 points on average ($P=.003$). For perspective, consider that participants scored 2,029 points on average in the one-cent piece-rate condition (i.e., when researchers paid them one cent per 100 points within a day of taking the study). When researchers delayed the payment by two weeks (i.e., researchers paid them one cent per 100 points two weeks after taking the study), participants scored 2,004 points. Therefore, the two-week delay decreased effort by 2,029-2,004=25 points. Economists predicted that the two-week delay would decrease effort by 71 points. Although the economists overestimated the impact of every psychological phenomenon, only three were statistically significant after correcting for multiple hypothesis testing: motivational crowding out, time discounting, and social preferences (see Figure \ref{fig:bias}).

\begin{figure}
    \centering
    \includegraphics[scale=.7]{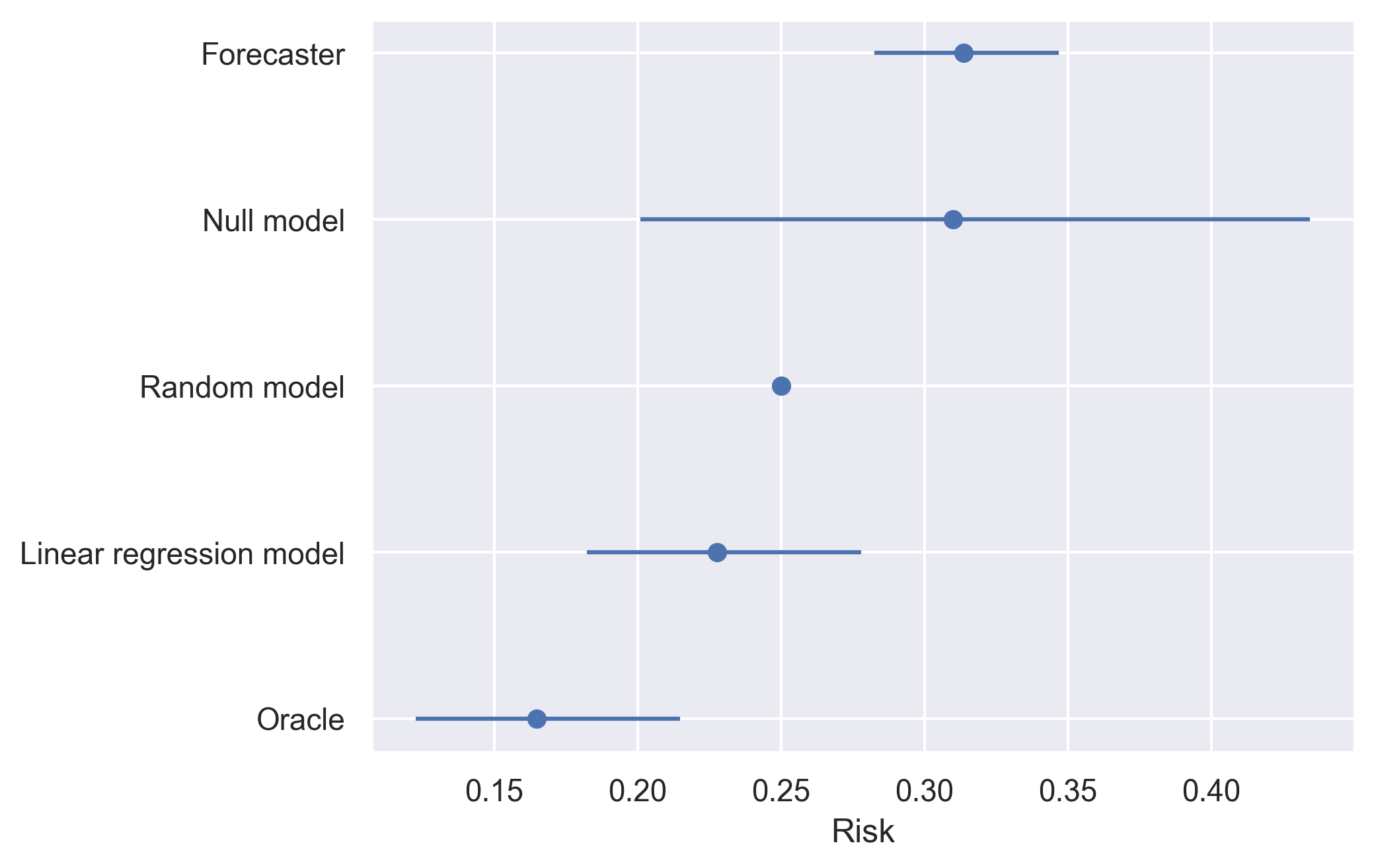}
    \caption{Risk for the reproducibility study.}
    \label{fig:risk}
\end{figure}

\textbf{Reproducibility study.} Figure \ref{fig:risk} shows that the psychologists' reproducibility study predictions were worse than all three models we considered: the null model, random chance, and linear regression. Even after correcting for multiple hypothesis testing, this difference was statistically significant for random chance ($M=.06, P < .001$) and linear regression ($M=.09, P=.014$). Psychologists estimated that the average published psychology study had a 54\% chance of replicating. The reproducibility study results suggest that the average study had only a 32\% chance of replicating. This bias was statistically significant ($P=.002$).

\section{Discussion}

Critical public policy decisions depend on predictions from behavioral scientists. In this paper, we asked how accurate those predictions are. To answer this question, we compared the predictions of 640 behavioral scientists to those of simple mathematical models on five prediction tasks. Our sample included a variety of behavioral scientists: economists, psychologists, and business professionals from academia, industry, and government. The prediction tasks also covered various domains, including text-message interventions to increase vaccination rates, behavioral nudges to increase exercise, randomized control trials, incentives to encourage effort, and attempts to reproduce published psychology studies. The models to which we compared the behavioral scientists were deliberately simple, such as random chance, linear interpolation, and heuristics like ``behavioral interventions have no effect" and ``all published psychology research is false."

We consistently found that behavioral scientists are no better than - and often worse than - these simple heuristics and models. In the exercise, flu, and RCT studies, null models significantly outperformed behavioral scientists. These null models assume that behavioral treatments have no effect; behavioral interventions will not increase weekly gym visits, text messages will not increase vaccination rates, and nudges will not change behavior. As we can see in Table \ref{table:risk}, compared to behavioral scientists, null models are nearly indistinguishable from the oracle.

In the effort study, linear interpolations performed at least as well as professional economists. These interpolations assumed that all psychological phenomena are inert; people do not exhibit risk aversion, time discounting, or biases like framing effects.

In the reproducibility study, professional psychologists' Brier scores were virtually identical to those of a null model, which assumed that all published psychology research is false. Professional psychologists were significantly worse than both linear regression and random chance.

Notably, the linear regression model used data from the reproducibility study, which were not accessible to psychologists during their participation. While this is not a fair comparison, we believe it is a useful comparison, as the linear regression model can serve as a benchmark for future attempts to predict reproducibility.

Why is it so hard for behavioral scientists to outperform simple models? One possible answer is that human predictions are noisy while model predictions are not \citep{kahneman2021noise}. Indeed, there is likely a selection bias in the prediction tasks we analyzed. Recall that most of the prediction tasks asked behavioral scientists to predict the results of ongoing or recently completed studies. Behavioral scientists presumably spend time researching questions that have not been studied exhaustively and do not have obvious answers. In this case, the prediction tasks were likely exceptionally challenging, and behavioral scientists' expertise would be of little use.

However, behavioral scientists' predictions are not only noisy but also biased. Previous research noted that behavioral scientists overestimate the effectiveness of nudges \citep{dellavigna2022rcts, milkman2021megastudies}. Our research extends these findings, suggesting that behavioral scientists believe behavioral science generally ``works" better than it does. Behavioral scientists overestimated the effectiveness of behavioral interventions in the exercise, flu, and RCT studies. In the exercise study, behavioral scientists significantly overestimated the effectiveness of all 53 treatments, even after correcting for multiple testing. Economists overestimated the impact of psychological phenomena in the effort study, especially for motivational crowding out, time discounting, and social preferences. Finally, psychologists significantly overestimated the replicability of published psychology research in the reproducibility study. In general, behavioral scientists overestimate not only the effect of nudges, but also the impact of psychological phenomena and the replicability of published behavioral science research.

Behavioral scientists' bias can have serious consequences. A recent study found that policymakers were less supportive of an effective climate change policy (carbon taxes) when a nudge solution was also available \citep{hagmann2019nudging}. However, accurately disclosing the nudge's impact shifted support back towards carbon taxes and away from the nudge solution. In general, when behavioral scientists exaggerate the effectiveness of their work, they may drain support and resources from potentially more impactful solutions.

Our results raise many additional questions. For example, is it only behavioral scientists who are biased, or do people, in general, overestimate how well behavioral science works? The general public likely has little exposure to RCTs, social science experiments, and academic psychology publications, so there is no reason to expect that they are biased in either direction. Then again, the little exposure they have had likely gives an inflated impression of behavioral science's effectiveness. For example, a TED talk with 64 million as of May 2022 touted the benefits of \textit{power posing}, whereby one can reap the benefits of improved self-confidence and become more likely to succeed in life by adopting a powerful pose for one minute \citep{carney2010power, cuddy2012power}. However, the power posing literature was based on p-hacked results \citep{simmons2017power}, and researchers have since found that power posing yields no tangible benefits \citep{jonas2017power}.

Additionally, people may generally overestimate effects due to the ``What you see is all there is" (WYSIATI) bias \citep{kahneman2011thinking}. For example, the exercise study asked behavioral scientists to consider, among other treatments, how much more people would exercise if researchers told them they were ``gritty." After the initial ``gritty diagnosis," dozens of other factors determined how often participants in that condition went to the gym during the following four-week intervention period. Work schedule, personal circumstances, diet, mood changes, weather, and many other factors also played key roles. These other factors may not have even crossed the behavioral scientists' minds. The WYSIATI bias may have caused them to focus on the treatment and ignore the noise of life that tempers the treatment's signal. Of course, this bias is likely to cause everyone, not only behavioral scientists, to overestimate the effectiveness of behavioral interventions and the impact of psychological phenomena.

If people generally overestimate how well behavioral science works, are they more or less biased than behavioral scientists? Experimental economics might suggest that behavioral scientists are less biased because people with experience tend to be less biased in their domain of expertise. For example, experienced sports card traders are less susceptible to the endowment effect \citep{list2004testing}, professional traders exhibit less ambiguity aversion than novices \citep{list2010investment}, experienced bidders are immune to the winner's curse \citep{harrison2008naturally}, and CEOs who regularly make high-stakes decisions are less susceptible to possibility and certainty effects \citep{list2011ceos}. Given that most people have zero experience with behavioral science, they should be more biased than behavioral scientists.

Then again, there are at least three reasons to believe that behavioral scientists should be more biased than the general population: selection bias, selective exposure, and motivated reasoning. First, behavioral science might select people who believe in its effectiveness. On the supply side, students who apply to study psychology for five years on a measly PhD stipend are unlikely to believe that most psychology publications fail to replicate. On the demand side, marketing departments and nudge units may be disinclined to hire applicants who believe their work is ineffective. Indeed, part of the experimental economics argument is that markets filter out people who make poor decisions \citep{list2008market}. The opposite may be true of behavioral science: the profession might filter out people with an accurate assessment of how well behavioral science works.

Second, behavioral scientists are selectively exposed to research that finds large and statistically significant effects. Behavioral science journals and conferences are more likely to accept papers with significant results. Therefore, most of the literature behavioral scientists read promotes the idea that behavioral interventions are effective and psychological phenomena substantially influence behavior. However, published behavioral science research often fails to replicate. Lack of reproducibility plagues not only behavioral science \citep{open2012open, open2015estimating, camerer2016evaluating, mac2022evaluating} but also medicine \citep{freedman2015economics, prinz2011believe}, neuroscience \citep{button2013power}, and genetics \citep{hewitt2012editorial, lawrence2013mutational}. Scientific results fail to reproduce for many reasons, including publication bias, p-hacking, and fraud \citep{simmons2011false, nelson2018psychology}. Indeed, most evidence that behavioral scientists overestimate how well behavioral science works involves asking them to predict the results of nudge studies. However, there is little to no evidence that nudges work after correcting for publication bias \citep{maier2022no}. Even when a study successfully replicates, the effect size in the replication study is often much smaller than that reported in the original publication \citep{camerer2016evaluating, open2015estimating}. For example, the RCT study paper estimates that the academic literature overstates nudges' effectiveness by a factor of six \citep{dellavigna2022rcts}.

Finally, behavioral scientists might be susceptible to motivated reasoning \citep{kunda1990case,epley2016mechanics}. As behavioral scientists, we want to believe that our work is meaningful, effective, and true. Motivated reasoning may also drive selective exposure \citep{benabou2002self}. We want to believe our work is effective, so we disproportionately read about behavioral science experiments that worked.

Our analysis finds mixed evidence of the relationship between experience and bias in behavioral science. The RCT study informally examined the relationship between experience and bias for behavioral scientists predicting nudge effects and concluded that more experienced scientists were less biased. While we also estimate that more experienced scientists are less biased, we do not find statistically significant pairwise differences between the novice, moderately experienced, and most experienced scientists.

Even if the experimental economics argument is correct that behavioral scientists are less biased than the general population, why are behavioral scientists biased at all? The experimental economics literature identifies two mechanisms to explain why more experienced people are less biased \citep{list2003does, list2008market}. First, markets filter out people who make poor decisions. Second, experience teaches people to think and act more rationally. We have already discussed that the first mechanism might not apply to behavioral science. And, while our results are consistent with the hypothesis that behavioral scientists learn from experience, they still suggest that even the most experienced behavioral scientists overestimate the effectiveness of nudges. The remaining bias for the most experienced scientists is larger than the gap between the most experienced scientists and novices. Why has experience not eliminated this bias entirely? Perhaps the effect of experience competes with the forces of ``What you see is all there is," selection bias, selective exposure, and motivated reasoning such that experience mitigates but does not eliminate bias in behavioral science.

Finally, how can behavioral scientists better forecast behavior? One promising avenue is to use techniques that help forecasters predict political events \citep{chang2016developing,mellers2014psychological}. For example, the best political forecasters begin with base rates and then adjust their predictions based on information specific to the event they are forecasting \citep{tetlock2016superforecasting}. Behavioral scientists' predictions would likely improve by starting with the default assumptions that behavioral interventions have no effect, psychological phenomena do not influence behavior, and published psychology research has a one in three chance of replicating \citep{open2012open}. Even though these assumptions are wrong, they are much less wrong than what behavioral scientists currently believe.

\section{Methods}

\label{section:methods}

\subsection{Setup}

All of the studies we considered have a similar data generating process. First, the researchers selected a set of $K$ treatments to test. Denote the true effects of these treatments as $\mu \coloneqq (\mu_1,...,\mu_K)^T$.

Then, the researchers asked $F$ behavioral scientists to predict how effective the selected treatments would be. Let $X$ be the $K \times F$ matrix of the behavioral scientists' predictions, where $X_{k,f}$ is behavioral scientist $f$'s prediction about the effect of treatment $k$. We also predict the treatment effects using a model and call these predictions $\mu^m \coloneqq (\mu^m_1,...,\mu^m_K)^T$.

Finally, the researchers ran the experiment, which gave them noisy estimates $Y \coloneqq (Y_1,...,Y_K)^T$ of the treatment effects. By the central limit theorem, these estimates are approximately normally distributed $Y \sim \mathcal{N}(\mu, \Sigma)$, where $\Sigma$ is the $K \times K$ covariance with which the treatment effects were estimated.

This setup above applies to the exercise, flu, RCT, and effort studies. Much of the same setup applies to the replication study as well. Instead of $K$ treatments to test, the researchers selected $K$ studies to replicate. $\mu_k$ is the true probability that study $k$ will replicate. The probability that study $k$ will replicate depends on the sample size of the replication study $n_k$ and the effect of the study's treatment $\mu^*_k$. Specifically, 

\[
    \mu_k = \Phi(\sqrt{n_k} \mu^*_k - c_\alpha)
\]

where $c_\alpha$ is the critical value for significance level $\alpha$ (around 1.96 for a two-tailed test with $\alpha=.05$). Note that we normalize $\mu^*_k$ by study $k$'s sample standard deviation for comparability across studies.

When the researchers ran the replication studies, they observed noisy estimates $Y^*$ of the treatment effects. By the central limit theorem, these estimates are approximately normally distributed $Y^* \sim \mathcal{N}(\mu^*, \Sigma^*)$, where $\Sigma^*$ is a $K \times K$ diagonal covariance matrix where the $k$th diagonal element is $1 / n_k$.

We can back out $Y^*_k$ from the replication study data set using the sample size $n_k$, p-value $p_k$, and the direction of the estimated effect.

\[
    \sqrt{n_k} Y^*_k = \begin{cases}
        \Phi^{-1}(1 - p_k / 2) & \text{if $Y^*_k > 0$} \\
        \Phi^{-1}(p_k / 2) & \text{else}
    \end{cases}
\]

\subsection{Bias}

Bias is how much behavioral scientists overestimate the effects of treatments in general,

\[
    B(\mu, X) \coloneqq \E_\mu\big[\E[X_{k,f} - \mu_k]\big].
\]

If the bias is positive, the behavioral scientists overestimated the treatment effects.

Similarly, the bias for a given treatment $k$ is

\[
    B_k(\mu, X) \coloneqq \E_\mu\big[\E[X_{k,f} - \mu_k | k]\big].
\]

This is how we measured the bias for seven of the eight categories of psychological phenomena tested in the effort study.

The bias for the difference between two treatments $k$ and $l$ is

\[
    B_{k,l}(\mu, X) \coloneqq \E_\mu\big[\E[(X_{k,f} - X_{l,f}) - (\mu_k - \mu_l) | k, l]\big].
\]

If the bias is positive, the behavioral scientists overestimated the difference between the effectiveness of treatments $k$ and $l$. This is how we measured bias for the framing effects category in the effort study. Specifically, $k$ was a loss-framed condition (the researchers gave participants \$0.40, which they would lose if they scored fewer than 2,000 points) and $l$ was a gain-framed condition (the researchers promised participants \$0.40 if they scored at least 2,000 points).

\subsection{Risk}

Loss functions measure how accurately forecasters and models predicted the results of the experiment. Let $l : \mathbb{R} \times \mathbb{R} \to \mathbb{R}$ be a generic loss function that takes the true and predicted treatment effect and returns a real value (the loss). For example, $l$ could be squared error. So, the squared error for behavioral scientist $f$'s prediction about the effect of treatment $k$ is $l(\mu_k, X_{f,k}) = (\mu_k - X_{f,k})^2$.

Behavioral scientists' corresponding risk function is the expected loss 

\[
    R(\mu, X) \coloneqq \E_\mu\big[\E[l(\mu_k, X_{k,f})]\big]
\]

If the risk is low, the behavioral scientists' predictions were accurate. If the risk is high, the predictions were inaccurate.

Our main outcome variable is the \textit{comparative risk}; the expected difference between the behavioral scientists' loss and a model's loss

\[
    CR(\mu, X, \mu^m) \coloneqq \E_\mu\big[\E[l(\mu_k, X_{k,f}) - l(\mu_k, \mu^m_k)]\big].
\]

If the comparative risk is negative, the behavioral scientists' predictions were better than those of the model. If the comparative risk is positive, the model's predictions were better than those of the behavioral scientists.

\subsection{Estimating bias and risk}

We now turn to our procedure for estimating bias and risk. It is tempting to simply replace the true effects $\mu$ with their noisy estimates $Y$ and then use standard statistical methods. For example, it is tempting to test whether behavioral scientists are biased for treatment $k$ by taking $X_{k,1} - Y_k,...,X_{k,F} - Y_k$ as our sample and testing whether the mean is different from 0 using a $t$-test.

However, this ``plug-in procedure" ignores the variability in $Y_k$ as an estimate of $\mu_k$. This could lead us to conclude that behavioral scientists are biased when in fact they are not. To see this, suppose that behavioral scientists are unbiased, $\E[X_f] = \mu$. By chance, $Y_k$ underestimates $\mu_k$. Then, as the number of behavioral scientists in our sample increases, the sample mean of $X_{1.k} - Y_k,...,X_{k,F} - Y_k$ will converge to $\mu_k - Y_k > 0$, making it appear as if behavioral scientists are biased. A similar problem applies to plugging in $Y_k$ for $\mu_k$ when estimating risk.

In sum, there are three sources of uncertainty:

\begin{enumerate}
    \item Researchers sampled treatments from a population of treatments they could have tested. For example, in the exercise study, this population is ``the sort of behavioral interventions researchers would use to try to increase exercise in an study like this."
    \item Researchers sampled behavioral scientists from a population of scientists they could have asked to predict the experimental results.
    \item Researchers obtained a noisy estimate of the true treatment effects.
\end{enumerate}

To account for our uncertainty about the true treatment effects, we use empirical Bayes estimators from the \verb|multiple-inference| statistics package \citep{Bowen2022multiple} to estimate the posterior distribution of $\mu | Y$. When possible, we used nonparametric empirical Bayes estimators to avoid making parametric assumptions about the prior distribution of treatment effects. Nonparametric empirical Bayes estimators perform well with many data points and usually assume that the treatment effects are estimated independently \citep{cai2021nonparametric, brown2009nonparametric}. Therefore, we used nonparametric empirical Bayes for the RCT and reproducibility studies. We used parametric (normal-prior, normal-likelihood) empirical Bayes for the exercise and flu studies because the treatment effects were estimated with correlated errors. We also used parametric empirical Bayes for the effort study because it included only 18 treatments.

To account for uncertainty about the populations of treatments and behavioral scientists, we use a Bayesian bootstrap with a uniform Dirichlet prior and sample weights at the treatment level and the at behavioral scientist level \citep{rubin1981bayesian}. Sampling weights at the treatment level accounts for the fact that some treatment effects are harder to predict than others. Sampling weights at the behavioral scientist level accounts for the fact that some behavioral scientists are better forecasters than others. This is the bootstrap equivalent of clustering standard errors by treatment and behavioral scientist.

For concreteness, Algorithm \ref{alg:comparative_risk} describes how to use our sampling procedure to estimate the probability that behavioral scientists' predictions are more accurate than those of a mathematical model. That is, we estimate the probability that the comparative risk is negative. We use similar sampling procedures to estimate bias.

\begin{algorithm}
\caption{Estimating comparative risk}
\label{alg:comparative_risk}
\begin{algorithmic}
\Require $K \times 1$ vector of estimated effects $Y$ and $K \times K$ covariance matrix $\Sigma$
\Require $K \times 1$ vector of model predictions $\mu^m$
\Require $K \times F$ matrix of behavioral scientist predictions $X$
\Require Number of bootstrap samples to draw $N_S$
\State $N_T \gets 0$
\State $P \gets$ posterior distribution of $\mu | Y$ using empirical Bayes
\For{$s = \{1,...,N_S\}$}
    \State Sample $\mu^s$ from $P$
    \State Sample $w$ from $\text{Dir}(\mathbf{1}_K)$ where $\mathbf{1}_K$ is a $K \times 1$ vector of 1's
    \State Sample $m$ from $\text{Dir}(\mathbf{1}_F)$
    \State Compute $L$ such that $L_{k,f} = l(\mu^s_k, X_{k,f}) - l(\mu^s_k, \mu^m_k)$
    \If{$w^T L m < 0$}
        \State $N_T \gets N_T + 1$
    \EndIf
\EndFor
\State $\hat{Pr}_\mu\big\{CR(\mu, X, \mu^m) < 0\big\} = N_T / N_S$
\end{algorithmic}
\end{algorithm}

\subsection{Linear regression model}

Here, we describe how we use linear regression to predict a study's probability of replicating using data from the Reproducibility Project \citep{open2012open}. The linear regression model itself is standard. Our feature matrix $Z$ is a $K \times 2$ matrix containing a constant regressor and the estimated effect in the original study (normalized by the sample standard deviation of the original study). We can compute point estimates simply by regressing $Y^*$ on $Z$ and predicting the effect of replication study $k$ as $\hat{Y}^*_k = \hat{\beta}^T Z_k$, where $\hat{\beta}$ is a vector of OLS coefficients. Then, we can estimate the replication probability as,

\[
	\hat{Pr}\{\text{replicate}_k\} = \Phi(\sqrt{n_k} \hat{Y}^*_k - c_\alpha)
\]

However, this procedure does not account for the uncertainty in $\hat{Y}^*_k$ as an estimate of $Y^*_k$. Algorithm \ref{alg:linear_regression} describes a Gibbs sampling procedure to estimate the probability that study $k$ will replicate using a linear regression model. Additionally, this algorithm assumes that the variance of the error term is a linear function of $Z$, $\sigma^2 = Z \gamma$.

\begin{algorithm}
\caption{Estimating replication probability using linear regression}
\label{alg:linear_regression}
\begin{algorithmic}
\Require $K \times 2$ feature matrix with a constant regressor and the estimated effect size from the original study
\Require $K \times 1$ vector of estimated effects $Y^*$ from the replication study
\Require Number of Gibbs samples to draw $N$
\State $P \gets \emptyset$
\For{$s = \{1,...,N\}$}
    \State Regress $Y^*$ on $Z$ to obtain a joint distribution of OLS coefficients $\hat{\beta}$
    \State Sample $\beta^S$ from the joint distribution of $\hat{\beta}$
    \State Regress $(Z \beta^s - Y^*)^2$ on $Z$ to obtain a joint distribution of OLS coefficients $\hat{\gamma}$
    \State Sample $\gamma^s$ from the joint distribution of $\hat{\gamma}$
    \State Sample $Y^s_k$ from $\mathcal{N}({\beta^s}^T Z_k, {\gamma^s}^T Z_k)$
    \State $P \gets P \cup \big\{\Phi(\sqrt{n_k} Y^s_k - c_\alpha)\big\}$
\EndFor
\State $\hat{Pr}\{\text{replicate}_k\} = \frac{1}{N} \sum_s P^s$
\end{algorithmic}
\end{algorithm}

\clearpage
\singlespacing
\bibliographystyle{unsrtnat}
\bibliography{bibliography}

\end{document}